\documentclass[a4paper]{article}

\usepackage{graphicx}

\renewcommand{\theequation}{\arabic{equation}}

\begin{document}

\renewcommand{\theequation}{\arabic{equation}}

\begin{center}
{\Large {\bf Self-Duality and Phase Structure of the\\
 4D Random-Plaquette $Z_2$ Gauge Model }}

\vskip 0.5cm
{\Large Gaku Arakawa\footnote{e-mail address:
e101608@phys.kyy.nitech.ac.jp} and Ikuo Ichinose\footnote{e-mail 
 address: ikuo@nitech.ac.jp}}  

\vskip 0.3cm

{Department of Applied Physics, 
Nagoya Institute of Technology,
Nagoya 466-8555, Japan}

\vskip 0.5cm

{\Large Tetsuo Matsui\footnote{e-mail address:
matsui@phys.kindai.ac.jp}}

\vskip 0.3cm

{Department of Physics, Kinki University, Higashi-Osaka 
577-8502, Japan}

\vskip 0.5cm

{\Large Koujin Takeda\footnote{e-mail address:
takeda@stat.phys.titech.ac.jp}}

\vskip 0.3cm

{Department of Physics, Tokyo Institute of Technology,
Oh-okayama, Meguro, Tokyo 152-8551, Japan}

\end{center}
\begin{center} 
\begin{bf}
Abstract
\end{bf}
\end{center}
In the present paper, we shall study the $4$-dimensional $Z_2$ lattice 
gauge model
with a random gauge coupling; the random-plaquette gauge model(RPGM).
The random gauge coupling at each plaquette 
takes the value $J$ with the probability $1-p$ and 
$-J$ with $p$.
This model exhibits a confinement-Higgs phase transition. 
We numerically obtain a phase boundary curve in the $(p-T)$-plane where 
$T$ is the ``temperature" measured in unit of $J/k_B$.
This model plays an important role in estimating the accuracy threshold of
a quantum memory of a toric code.
In this paper, we are mainly interested in
its ``self-duality" aspect, and the relationship with the 
random-bond Ising model(RBIM) in $2$-dimensions.
The ``self-duality" argument can be applied both for
RPGM and RBIM, giving the same duality equations, hence
predicting the same phase boundary. 
The phase boundary curve obtained by our numerical simulation 
almost coincides with this predicted phase boundary at the high-temperature
region.
The phase transition is of first order for relatively small values of 
$p < 0.08$, but becomes of second order for larger $p$. 
The value of $p$ at the intersection of 
the phase boundary curve and the  Nishimori line is regarded as the 
accuracy threshold of errors in a toric quantum memory. 
It is estimated as $p=0.110\pm0.002$, which is very close to
the value conjectured by Takeda and Nishimori through the 
``self-duality" argument.

\newpage
\setcounter{footnote}{0}
\section{Introduction}

The duality transformations provide us with important informations
for certain class of statistical- and field-theoretical models.
The most famous one is the Kramers-Wannier self-duality relation 
for the $2$-dimensional(2D) Ising spin model\cite{KW}, which 
predicts the exact value of the critical temperature.

Its gauge-model counterpart was studied by Wegner\cite{weg,wu}, and
he  showed that the $4$-dimensional Ising($Z_2$) lattice gauge theory 
is also self-dual.
Its self-duality condition for the critical coupling constant
is equivalent to that of the 2D Ising spin model,
so the value of the critical coupling(temperature) coincides
with that of the Ising model, though the orders of the phase 
transitions may be (and actually are) different in 
these two systems\cite{IGT}.

Recently, a duality transformation is applied for a {\em random spin} 
system of spin glass, the random-bond Ising model(RBIM) in two 
dimensions\cite{Nishimori-let,MNN}.
This model contains a new parameter $p$, which controls the random 
quenched variables, i.e., 
the rate of nearest-neighbor spin-coupling 
with ``wrong sign". Thus the critical temperature  
becomes a function $T(p)$ of $p$.
The ``self-duality condition" for this model is proposed
in order to locate the multicritical point\cite{MNN}.
The multicritical point is the intersection of the critical temperature
$T(p)$ 
and the Nishimori line of spin glass. 
(More detailed discussion on this point will be given in Sec.2.)
The duality transformation for the 2D RBIM is
exact, but contrary to the nonrandom case ($p=0$),
the proposed ``self-duality condition" does {\em not} assure us
that the singular point of the free energy is located at 
the ``self-dual point".
However, once this ``self-duality condition"  is accepted, it 
allows us to make a conjecture on the location 
of $T(p)$.
After that, this conjecture for RBIM has been verified by numerical 
simulation\cite{RBI} for high-temperature region.

The 4-dimensional (4D) $Z_2$ random-plaquette gauge model(RPGM)
plays an important role in the theory of quantum memory.
Here the randomness $p$ is the probability that
the gauge coupling for each plaquette takes the ``wrong-sign".
(See Sect.2 for more details.) 
This model is used to predict  
the accuracy threshold of a 3-dimensional(3D) toric quantum 
code\cite{QM,QM1,QM2}.
Actually, the accuracy threshold $p_c$ of the errors of a 3D toric 
quantum memory is determined by the multicritical point
of the 4D RPGM\cite{TN}. 
The accuracy threshold of a 3D toric code $p_c$ is expected 
to be higher than that of a 2D toric code, $p_c = 0.033$\cite{OAIM}.  
Thus it is quite important to obtain the phase boundary
of the 4D RPGM.

Takeda and Nishimori\cite{TN,TN2} 
applied the  duality transformation for the 4D $Z_2$ RPGM, and 
assumed  the ``self-duality condition"  
in order to locate  the multicritical point as $p_c = 0.110028...$. 
The ``self-duality condition" in the 4D $Z_2$ RPGM
is the same with that of the 2D RBIM, and 
so the phase boundaries $T(p)$ of these two models are the same
if the ``self-duality condition" correctly predicts the phase
boundary.

In this paper, we shall study the phase structure of the 4D RPGM
by numerical simulation, in particular its confinement-Higgs
phase transition. The result of the phase boundary curve
 shall be used to determine the accuracy threshold of a 3D toric code 
as well as to judge whether  
the ``self-duality condition" proposed to determine
$T(p)$ in  Ref.\cite{TN,TN2} is valid.
We calculate the phase boundary curve, i.e., 
the critical temperature $T(p)$, in the $p-T$ plane. 
The curve $T(p)$ starts from the critical point of 
the {\em nonrandom} gauge model with the uniform
coupling constant at $p=0$ and decreases as the randomness $p$ increases.
Our result $T(p)$ is plotted in Fig.1 together with the
result of ``self-duality condition" of Ref.\cite{TN,TN2} [See Eq.(\ref{pK})].
The order of the phase transition is of first order for $p<0.08$,
 but it becomes of second order at larger values of $p$.
We recall that the similar  behavior in the order of phase transition
is observed in the 3D RPGM, in which 
the phase transition changes from the second order transitions
to the higher order ones\cite{OAIM}.
Our curve  $T(p)$ of the 4D RPGM   
coincides with the predicted phase boundary by the ``self-duality
conditions" of the both models mentioned above.
This result verifies the conjecture by Takeda and Nishimori\cite{TN,TN2} 
for the multicritical point of the 4D RPGM.

This paper is organized as follows.
In Sec.2, we briefly review the duality transformation of the random models
and show how the ``self-duality condition" determines the critical curve 
in the $p-T$ plane.
In Sec.3, we report our numerical calculations presenting
the specific heat, the expectation values of the Wilson loops,
and the fluctuations of the internal energy of the 4D RPGM, 
to determine the critical curve. 
Section 4 is devoted for conclusion.  

\section{Duality in the 4D random-plaquette gauge model}

In this section we shall briefly review the duality transformation
of the 4D RPGM following Ref.\cite{TN}.
We consider the 4-dimensional hypercubic lattice,
and put a $Z_2$ gauge variable $U_{x\mu} =\pm1$ on each link
$(x\mu)$, where $x$ denotes the site and $\mu=1,2,3,4$ is the 
index for four positive directions.
The energy $H$ of the model is given by 
\begin{equation}
H(U;\tau)=-J\sum_{P}\tau_P \prod_{P}U,
\label{H}
\end{equation}
where $J (> 0)$ is the strength of the (inverse) gauge coupling,  
$\prod_{P}U$ is the 
product of the four gauge variables $U_{x,\mu}$  
on the four links surrounding the plaquette $P$.
$\tau_P$ is the random variable for each plaquette $P$ 
taking $\tau_P = 1$ with the probability $1-p$ and
the ``wrong-sign" $\tau_P = -1$ with the probability $p\ (\in [0,1])$.

We employ the replica technique for taking the ensemble
averages over $\tau_P$.
For the $n$-replica system, the averaged partition function $Z_n$
is given as 
\begin{eqnarray}
&& Z_n=\langle Z^n(\tau) \rangle_{\mbox{\footnotesize{ens}}},\nonumber \\
&& Z^n(\tau)\equiv\Big(\prod_{\alpha=1}^n \sum_{U^\alpha_{x\mu}=\pm 1}\Big)
 \exp \Big(-\beta\sum_{\alpha=1}^n H(U^\alpha;\tau)\Big),\ 
 \beta\equiv\frac{1}{JT},
\label{Zn}
\end{eqnarray}
where $\alpha=1,\cdots, n$ is the replica index, and 
$\langle O(\tau) \rangle_{\mbox{\footnotesize{ens}}}$ denotes
an ensemble average over $\{\tau_P\}$.
The free energy $F$ is obtained by taking the limit $n\rightarrow 0$ as usual,
\begin{eqnarray}
F &=& -\frac{1}{\beta}\lim_{n \rightarrow 0}\left( \frac{Z_n -1}{n}
\right).
\end{eqnarray}

Let us introduce the Boltzmann weight for a single plaquette $P$,
$\kappa_+=\exp(K)$
for the fluxless configurations $\prod_P U=1$ 
and the weight $\kappa_-=\exp(-K)$ for the fluxfull ones, 
where
$K \equiv\beta J = 1/T$.
As we are considering $n$ replicas, it is useful to introduce 
the {\em generalized Boltzmann weight} for a single plaquette, i.e.,
for the configurations with fluxless $n-k$ replicas and fluxfull $k$ 
replicas, we assign the weight $x_k$ as\cite{MNN,TN} 
\begin{equation}
x_k=(1-p)\kappa_+^{n-k}\kappa_-^k+p\kappa_+^k\kappa_-^{n-k}.
\label{xk}
\end{equation}
Then the averaged partition function is expressed 
in terms of these weights, $x_1, \cdots, x_n$,
\begin{equation}
Z_n=Z_n(x_0,x_1,\cdots,x_n).
\label{Zn2}
\end{equation}

To make a duality transformation, we introduce
 the dual Boltzmann weights $x^\ast_k\ (k=0,\cdots,n)$ 
by the following discrete
Fourier transformations;
\begin{eqnarray}
x^\ast_{2m}&=&{1\over 2^{n/2}}(\kappa_++\kappa-)^{n-2m}
             (\kappa_+-\kappa_-)^{2m},   \nonumber  \\
x^\ast_{2m+1}&=&{1\over 2^{n/2}}(1-2p)(\kappa_++\kappa-)^{n-2m-1}
              (\kappa_+-\kappa_-)^{2m+1}.
\label{xkast}
\end{eqnarray}
Then the following duality relation can be derived\cite{TN},
\begin{equation}
Z_n(x_0,x_1,\cdots,x_n)=Z_n(x^\ast_0,x^\ast_1,\cdots,x^\ast_n),
\label{dual}
\end{equation}
up to an irrelevant overall constant.

In the standard (nonrandom) $Z_2$ gauge model with $p=0$, 
it is known that  a confinement-Higgs phase transition takes place 
at $K=K_C$ where $e^{-2K_C}=\sqrt{2}-1$.
This critical value is obtained by imposing
the self-duality condition $x_0=x_0^*$, with which the other $n$ 
conditions $x_k=x_k^* \; (k=1,\cdots, n)$ also hold (automatically) at $p=0$.
Then it is expected that the phase boundary of 
the confinement-Higgs phase transition evolves starting 
at $(p=0, T=1/K_C)$ into the region $0 < p$.
It is very interesting to see if  the ``self-duality condition", 
\begin{equation}
 x_0=x_0^\ast
\label{x=xast}
\end{equation}
predicts not only the location of the multicritical point 
but also that of the {\em whole phase boundary}.
Here we note that one cannot impose $n$ self-duality
conditions $x_k =x_k^*\ (k=1,\cdots,n)$ simultaneously for $0 < p$;
these equations are overcomplete and have no solutions
in contrast with the case of $p=0$.
The arguments of the original and transformed partition functions
are {\em not} equal even at the ``self-dual point".
Therefore it is {\em not} necessarily assured that  the ``self-duality
condition" (\ref{x=xast}) determines the singular point of $Z_n$.

Explicitly
in the limit $n\rightarrow 0$, Eq.(\ref{x=xast}) reduces to the following 
equation,
\begin{equation}
p=-{1\over 2K}\log \Big({1+e^{-2K} \over \sqrt{2}}\Big),
\label{pK}
\end{equation}
which determines the phase boundary curve $T(p)$ in the $p-T$ plane.
The multicritical point is defined as the intersection point of 
the two curves, the phase boundary $T(p)$ 
and the Nishimori line defined by 
\begin{eqnarray}
\exp(-2K)=\frac{p}{1-p}.
\label{nishimori-line}
\end{eqnarray}
The value $p=p_C$ at the multicritical point is regarded as the 
accuracy threshold for the error rate of a quantum memory
of the 3D toric code\cite{TN,OAIM}.
By using Eq.(\ref{pK}), Takeda and Nishimori\cite{TN}
determined $p_C$ as $p_C=0.110028...$, which is considerably 
larger than the accuracy threshold $p_c = 0.033$
for a 2D toric code\cite{OAIM}, as it is naturally expected.

Here we comment on the 2D RBIM.
In Ref.\cite{Nishimori-let,MNN},  
the duality transformation has been applied for the 2D RBIM. 
By imposing the same ``self-duality condition" as Eq.(\ref{x=xast}), 
one obtains just Eq.(\ref{pK}),  so these two models are predicted 
to have the same phase boundaries $T(p)$. 
The numerical simulation\cite{RBI2,RBI} of the RBIM
gives the phase boundary that almost coincides with
Eq.(\ref{pK}) in the region of the $p-T$ plane {\em above the Nishimori line}
(the {\em high}-$T$ {\em region}).\footnote{To obtain definite results by the
numerical studies for the {\em low}$-T$ region is rather difficult and 
it requires a very large number of random samples.}
As explained above, the ``self-duality condition" for the random
systems is just a conjecture, so 
it is interesting to see if it is satisfied 
also in the random gauge systems.
The results of the 4D RPGM will be reported in the following section. 


\section{Numerical study of the 4D $Z_2$ RPGM}

In this section, we shall show our results of numerical simulation 
for the phase structure of the 4D $Z_2$ RPGM, particularly 
whether the phase 
boundary of the confinement-Higgs phase transition coincides with
the predicted curve (\ref{pK}).
As explained in the introduction, there exists a confinement-Higgs
phase transition in the nonrandom gauge theory with $p=0$, and 
the critical coupling is given as $K_C=-{1\over 2} \ln (\sqrt{2}-1)$
by the self-duality nature of the model.
We expect that the phase boundary curve 
evolves from the point $(p=0,T=1/K_C)$ 
downward as $p$ increases in the $p-T$ plane since
the inclusion of random couplings tends to put the system in
a disordered phase.

In our simulation, we first generate  $\tau_P$ over the lattice 
randomly to prepare a sample. Then we perform Monte Carlo simulation
of this sample with Metropolis algorithm.
After repeating this procedure, we get the results of a set of samples.
Finally we average these results over the samples. 
We calculated the following quantities;
\begin{enumerate}
\item Internal energy and specific heat
\item Expectation values of the Wilson loop and their deviations from the
perimeter/area law
\item Fluctuations of internal energy and specific heat among samples
\end{enumerate}

\noindent
The internal energy and the specific heat 
are useful to determine the phase boundary and the 
order of the phase transition in the high-$T$ region.
As we explained before, the high-$T$ region is the region above the 
Nishimori line in the $p-T$ plane.
The second quantity, the Wilson loop, is an
order parameter of the gauge theory\cite{Wilson}; it obeys the area law 
in the confinement phase, while it obeys the perimeter law in the 
deconfinement phase.
In the present case, we use it to locate 
the phase boundary close to the Nishimori line in the high-$T$ region.
The third quantity is used to identify the multicritical point,
the intersection 
of the phase boundary curve and the Nishimori line.
As it can be proved exactly, the specific heat (and the internal energy)
shows {\em no} singular behavior on the Nishimori line\cite{nishimori,OAIM}.
Then it is rather difficult to identify the phase transition point near
the Nishimori line.
We use all the above three quantities to identify the transition points.   

In Fig.2, we plot the internal energy $\langle E \rangle$ 
and the specific heat $C$ per site at $p=0, 0.02,
0.04, 0.06, 0.08$ and $0.10$.
It seems obvious that the phase transition is of
first order for smaller value of $p$, $p<0.08$, whereas it becomes
of second order for larger $p$.
The critical points $T(p)$ in Fig.1 in the high-$T$ region are determined
from these peaks of $C$.
They are fairly in good agreements with the self-duality
prediction (\ref{pK}), but there still exist small differences.
We think  that they are due to the finite-size effect.
To check this point,  we study the size-dependence of the specific heat.
Fig.3 shows $C$ at $p = 0.08$ for the lattice sizes $N^4$ with $N= 6,12,16$.
We observe that the peak of $C$ becomes sharper and higher for
larger lattices, which verifies the second-order phase transition 
at $p =0.08$. The value of $T$ at the peak decreases gradually as $N$
increases. 
In Fig.4 the location of $T$ at the peak of $C$ is plotted versus the inverse
of the linear size $N$ of the lattice. As $N$ increases,
$T(p)$ gets closer to the predicted value $T(p= 0.08)=1.552...$ by the 
self-duality condition (\ref{pK}).

Next let us see the expectation value of the 
Wilson loop (for the $Z_2$ gauge theory it was first introduced by 
Wegner \cite{weg}), 
$W(C)$, for a close loop $C$ on the lattice,
 which is defined as follows;
\begin{eqnarray}
W(C)&=&\langle W(C,\tau)\rangle_{\mbox{\footnotesize{ens}}},
\nonumber\\
W(C,\tau)
&=&\prod_{x\mu}\sum_{U_{x\mu}} \Big( \prod_C U_{x,\mu}\Big)
\exp(-\beta H(U;\tau))/Z(\tau),
\nonumber\\
 Z(\tau)&=&\prod_{x\mu}\sum_{U_{x\mu}} \exp(-\beta H(U;\tau)).
\label{WL}
\end{eqnarray}
In the confinement phase of the gauge theory, $W(C)$ obeys
what is called the area law,
\begin{equation}
W(C) \propto e^{-\alpha A(C)},
\label{area}
\end{equation}
where $\alpha$ is a constant named string tension
and $A(C)$ is the minimum value of the area of a membrane that
covers the closed loop $C$.
On the other hand, in the deconfinement phase, 
$W(C)$ obeys the perimeter law,
\begin{equation}
W(C) \propto e^{-\gamma P(C)},
\label{peri}
\end{equation} 
where $\gamma$ is another constant and $P(C)$ is the perimeter of $C$.

We calculate $W(C)$ for various $p$'s with a fixed $K=1/T$
and fit the data by the area and perimeter laws.
In Fig.5, the typical results near the multicritical point
are shown.
These results clearly show that $W(C)$ changes its behavior 
from the area law to the perimeter law as $p$ increases.
In Fig.6, we present the results of $\chi^2$ fittings 
for $K=0.9$ and $1.0$.
We can (roughly) estimate the phase transition point by using these
results.

Finally, we study the fluctuations of 
$\langle E \rangle$ and $C$ over the samples. 
In the previous studies on the 2D RBIM and the 3D RPGM,
it was observed that the fluctuations of these quantities 
indicate signals of the phase transition\cite{energy,OAIM}.
In Fig.7, we present these fluctuations for $K=0.9$ and $1.0$.
The fluctuation of $C$ stays almost constant at large  $p$
(i.e., in the confinement phase), and it
increases considerably as $p$ decreases and passes a certain critical value.
The critical point of $p$ is very close to the phase transition point
 estimated by the other numerical calculations given above.
More precisely, we estimate the values of $p$ at the criticality
for $K=1.0$ as $p=0.110\pm 0.002$. 
Since this point is very close to
the expected multicritical point as seen in Fig.1, we think that it is
reasonable to use this value as the estimation of $p_c$
at the multicritical point.

\section{Conclusion}

In this paper we studied the 
4D RPGM numerically.
All the calculations consistently indicate the existence
of the confinement-Higgs phase transition continuing
from the nonrandom case $p=0$ to the random case $0 < p$.
The critical curve $T(p)$ is determined synthetically
from $\langle E \rangle$, $C$, $W(C)$, and  the fluctuations of
$\langle E \rangle$ and  $C$ over the samples.
We summarize the results in Fig.1, which are in good agreement
with the predicted value by the ``self-duality condition" (\ref{pK}).
In particular, our estimation of the  multicritical point 
is  $p_c=0.110\pm 0.002$, while the ``self-duality condition"
predicts  $p=0.110228..$. We regard the small but non-negligible
differences between the numerical result of $T(p)$
and (\ref{pK}) as the finite-size effect.
Our studies strongly indicate the correctness of the conjecture
by the ``self-duality" (\ref{x=xast}) not only for the 
spin glass model but also for the random gauge model in the high-$T$
region. Numerical simulations of the random models at low-$T$ region
require considerably many  samples to obtain  
reliable ensemble averages\cite{OAIM}, 
so we did not present $T(p)$ at the low-$T$ region
below the Nishimori line in Fig.1. 
It is reserved for the future problem.

\vspace{0.5cm}
\begin{center}
{\bf Acknowledgment}
\end{center}

One of the authors (K.T.) would like to thank H. Nishimori, Y. Ozeki and
T. Sasamoto for their suggestions and useful comments.
He was supported by the Grant-in-Aid for Scientific Research
on Priority Area ``Statistical-Mechanical Approach to
Probabilistic Information Processing'' and the 21st Century
COE Program at Tokyo Institute of Technology ``Nanometer-Scale
Quantum Physics''.



\begin{figure}[ht]
 \begin{center}
  \includegraphics[width=.8\linewidth]{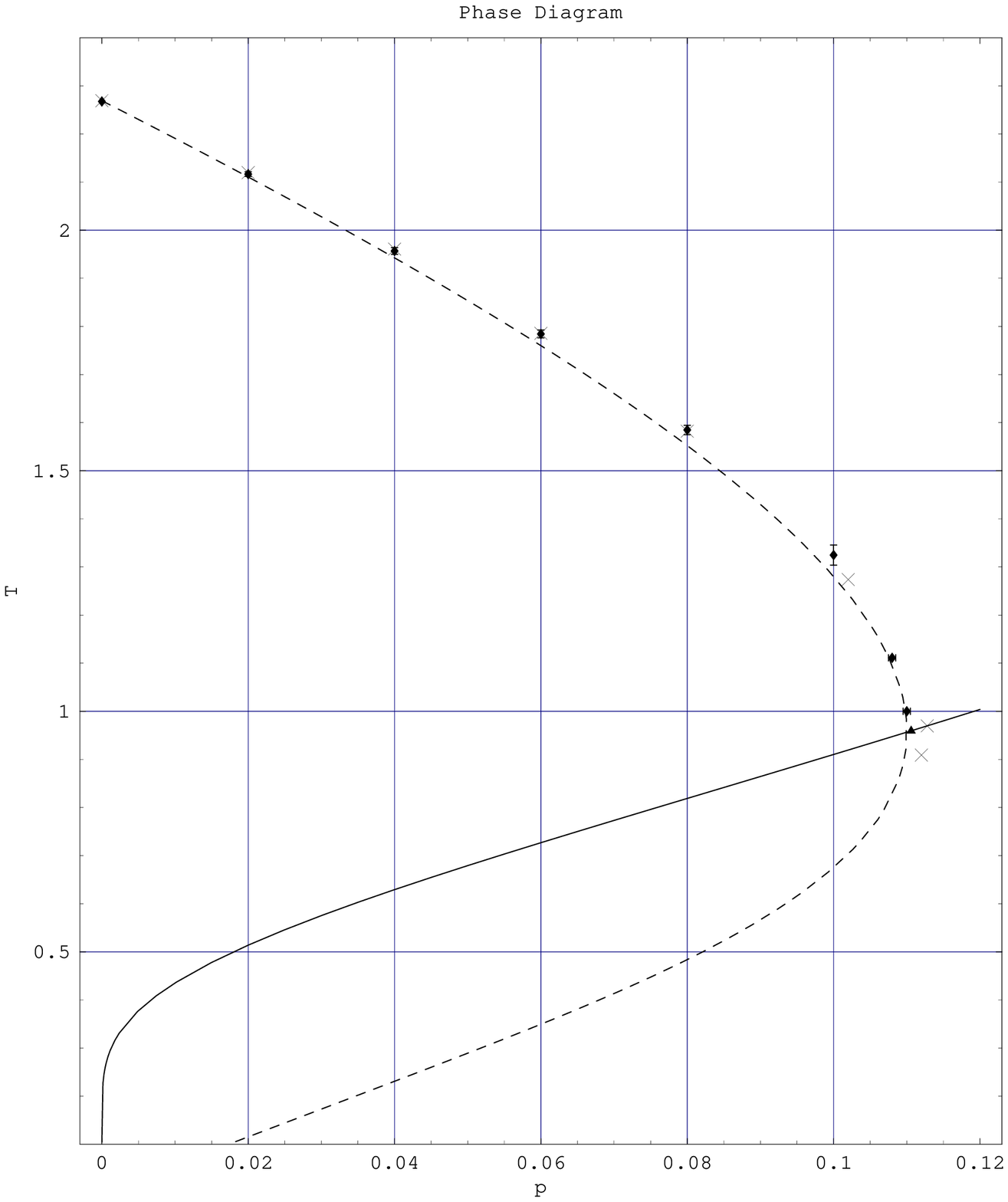} 
  \caption{Phase diagram of the 4D RPGM and the 2D RBIM 
  in the $p$(randomness)-$T$(temperature)
  plane. The diamond symbols show the transition points
  of the 4D RPGM obtained by our numerical study.  
The cross symbols show the transition points of the 2D RBIM 
calculated in Ref.\cite{RBI2}.
The triangle symbol shows the transition point of the 2D RBIM 
on the Nishimori line calculated in Ref.\cite{RBI}.
The dashed curve is the phase boundary obtained by the ``self-duality 
condition" (\ref{pK}), which predicts the same curve for the 4D RPGM 
and the 2D RBIM. This curve separate the confinement phase
on the right-hand side (larger $p$) and the Higgs(deconfinement) phase
on the left-hand side (smaller $p$). 
The solid curve  is the Nishimori line (\ref{nishimori-line}), 
which is also identical for the two models.}  
 \end{center}
\end{figure}

\begin{figure}[p]
 \begin{center}
  \includegraphics[width=.87\linewidth]{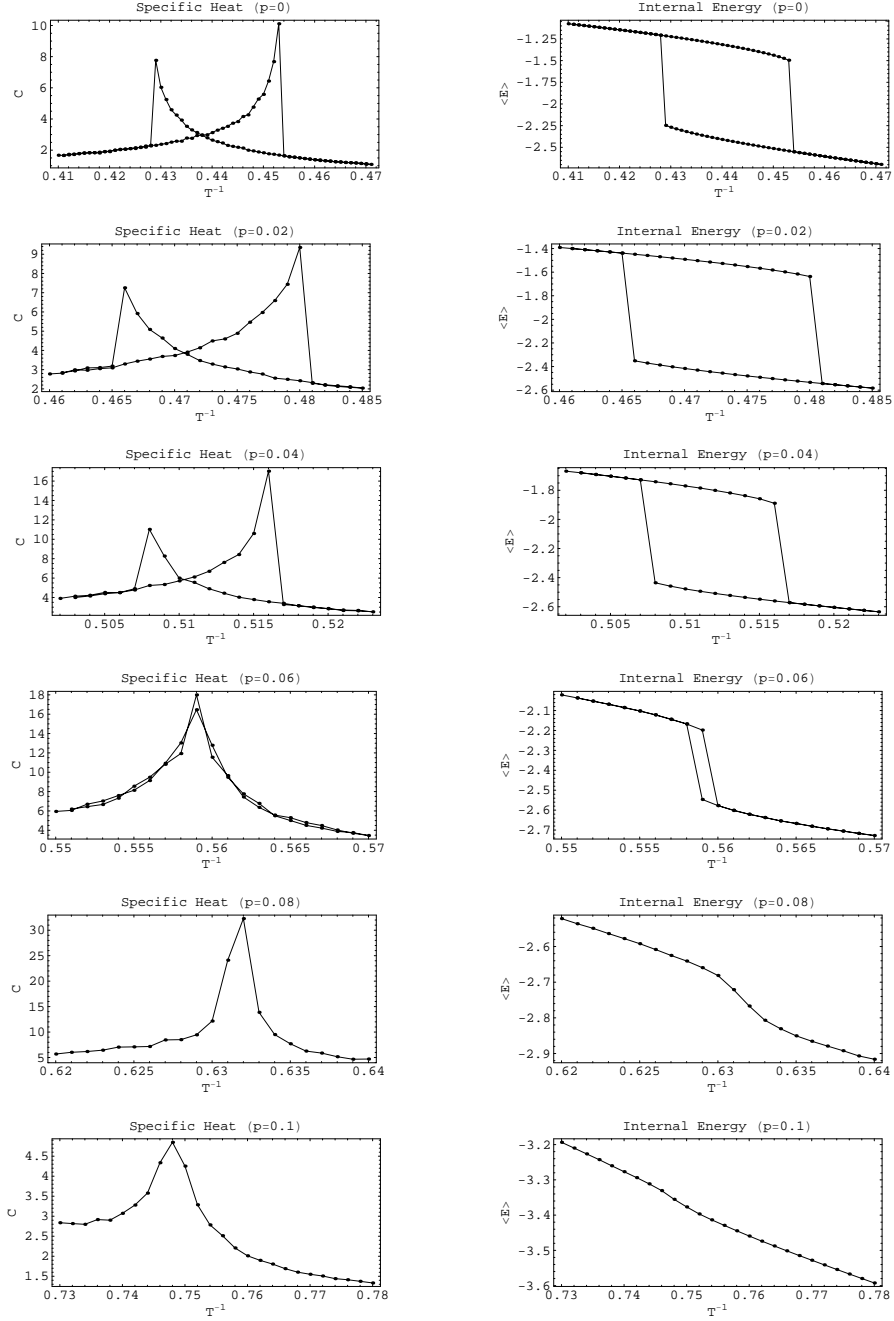} 
  \caption{Specific heat $C$ and internal energy $\langle E \rangle$
  per site  vs $1/T$ in the high-$T$ region for typical samples
  (from $p=0$ up to $p=0.1$).
  For $p<0.08$, $C$ exhibits a double-peak structure, and
  $\langle E \rangle$ exhibits hysteresis, both indicating that 
  the phase transition is of first order. The lattice size is
  $16^4$ for $p=0.0$ and $12^4$ for the other $p$'s, and the typical sweep
  number is $3\times 10^4$ for $p=0$ and $5\times 10^4(p=2,4,6\%), 10^5(p=8\%),
  2\times 10^6(p=10\%)$.}
 \end{center}
\end{figure}

\begin{figure}[ht]
 \begin{center}
  \includegraphics[width=.7\linewidth]{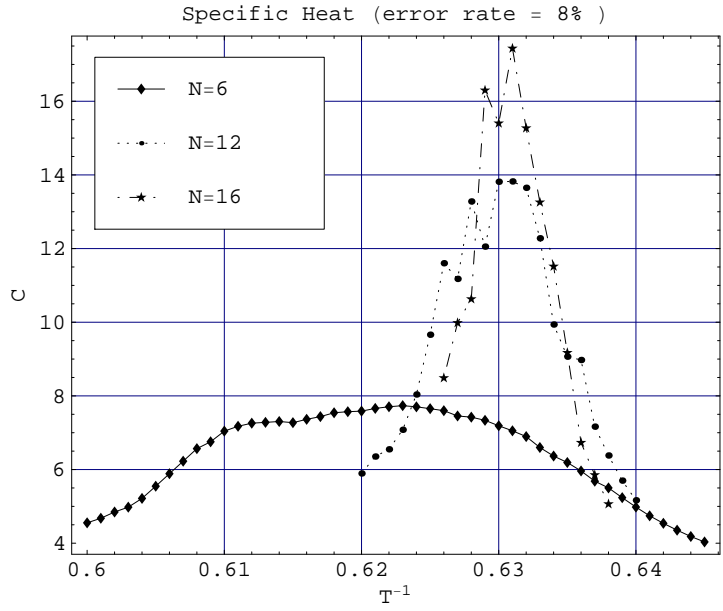}  
 \end{center}
 \caption{Lattice size dependence of the specific heat for $p=0.08$.
The number of used samples is $\sim 10^2$.}
\end{figure} 

\vspace{2cm}

\begin{figure}[ht]
 \begin{center}
  \includegraphics[width=.7\linewidth]{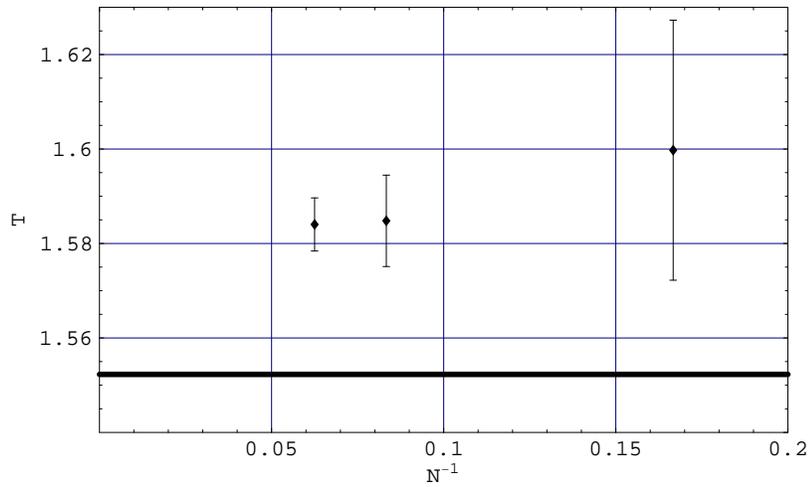} 
 \end{center}
 \caption{Lattice size dependence of the transition point for $p=0.08$. 
 The bold line indicates the predicted value $T=1.552...$ by 
the ``self-duality condition" (\ref{pK}).}
\end{figure}

\begin{figure}[ht]
 \begin{minipage}[c]{.12\linewidth}
  \setlength{\baselineskip}{4.5cm}
  $T^{-1}=0.9$ \\
  $T^{-1}=1.0$
 \end{minipage}
 \begin{minipage}[c]{.88\linewidth}
  \begin{center}
   \includegraphics[width=\linewidth]{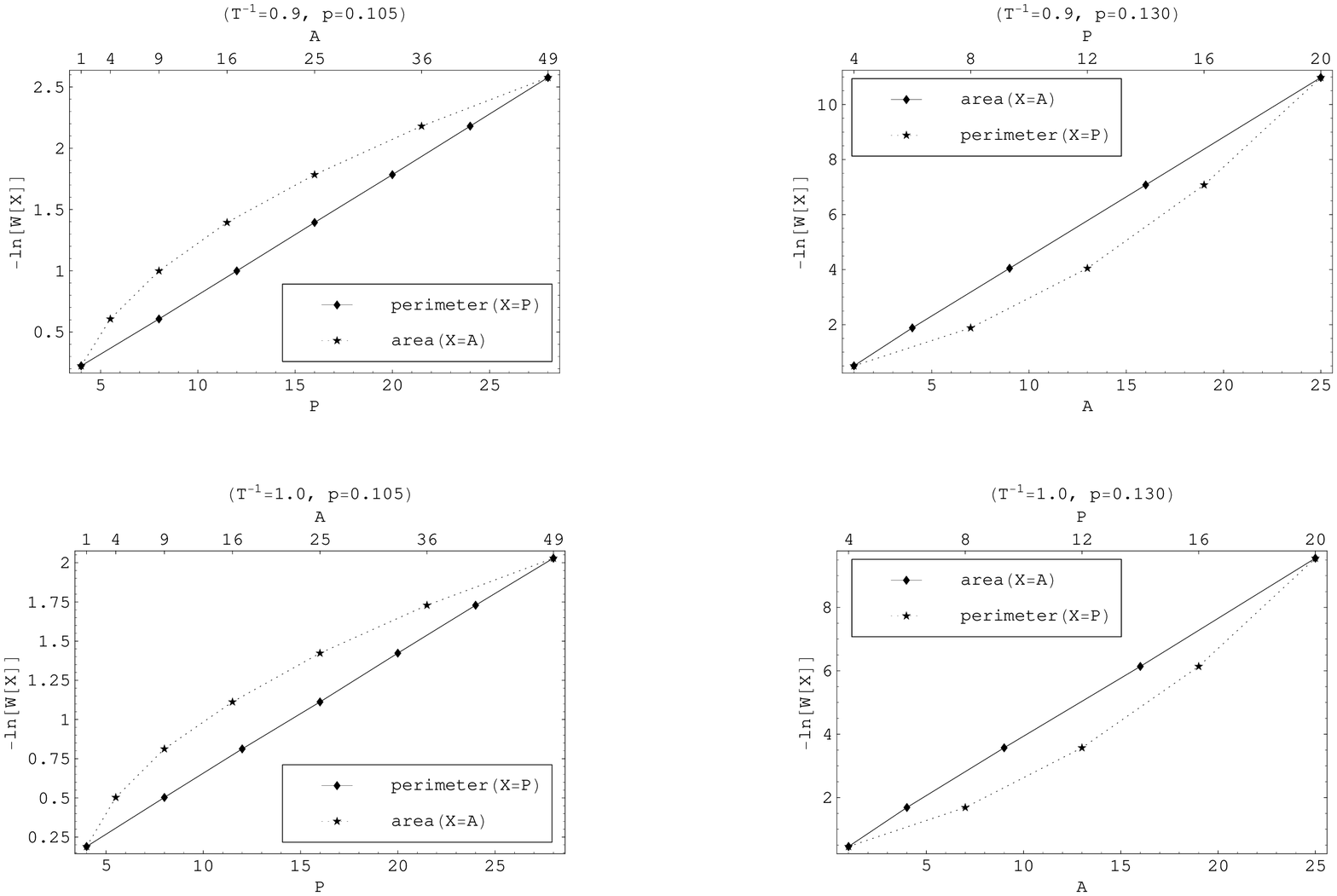}
   $p=0.105$ \hspace{5cm} $p=0.13$
  \end{center}
 \end{minipage}
 \caption{Wilson loops $W(C)$ for typical samples  
 in the region near the Nishimori line. 
The lower(upper) horizontal axis $P$ denotes the perimeter of the loop $C$ 
for the perimeter-law fitting, and the upper(lower) horizontal axis
$A$ denotes the area of C 
for the area-law fitting. The data show
 the perimeter(area)-law behavior
 for $p=0.105$($p=0.13$) in the left(right) column at each
 temperature. 
 The lattice size is $10^4$, and the typical 
 number of sweeps is $2\times10^4$.}
\end{figure}

\begin{figure}[ht]
 \begin{center}
  \includegraphics[width=\linewidth]{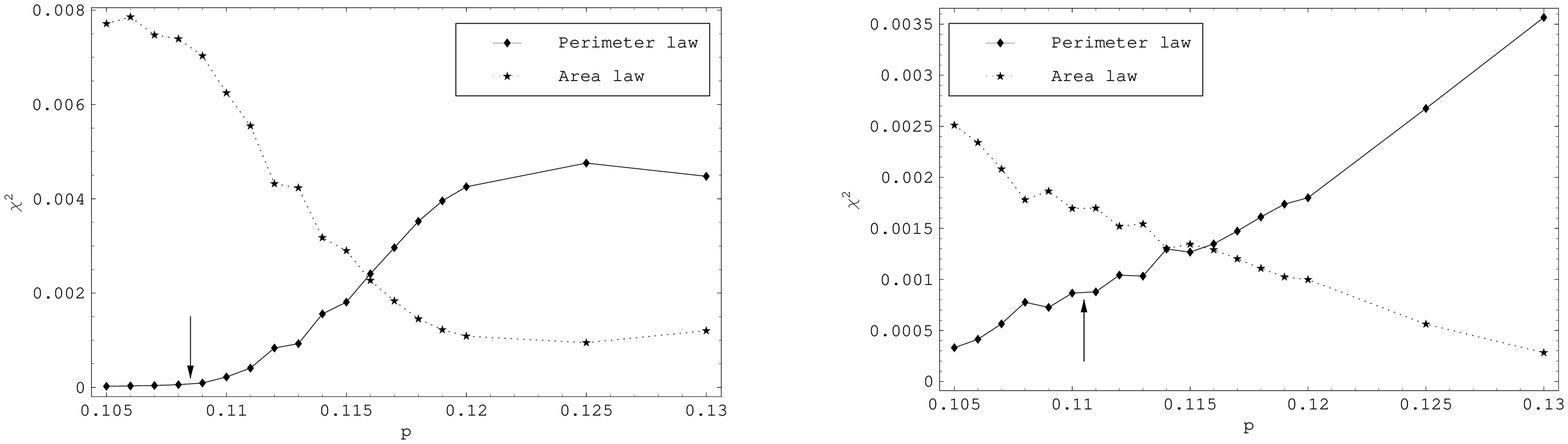}
  (a) $T^{-1}=0.9$ \hspace{5cm} (b) $T^{-1}=1.0$
  \caption{Deviation $\chi^2$ over 500 samples
  from the perimeter or area law vs $p$. 
  $W(C)$ is normalized as $-\ln[W(C_5)]=1$, where $C_5$ is the contour of
  a square whose area is $A(C_5)=5^2$. The arrows show the transition 
  points estimated from the specific heat and internal energy (see Fig.7).} 
 \end{center}
\end{figure}

\begin{figure}[ht]
 \begin{center}
  \includegraphics[width=\linewidth]{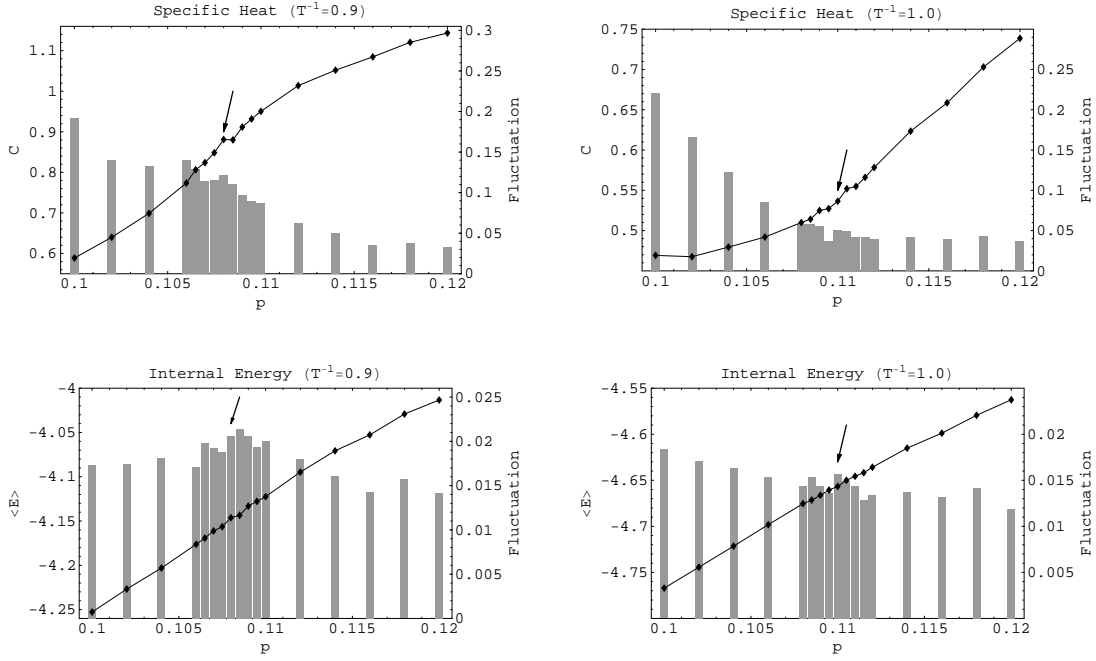}
  (a) $T^{-1}=0.9$ \hspace{5cm} (b) $T^{-1}=1.0$
 \end{center}
 \caption{Specific heat and internal energy in the region near
the Nishimori line. The histograms
show their  fluctuations among 200 samples.
The arrows exhibit the phase transition points at which the 
magnitudes of fluctuations change their rate of variation
with respect to $p$.
The lattice size is $10^4$, and the typical number of sweeps is
$2\times10^5$.}
\end{figure}

\end{document}